\documentclass[12pt]{article}
\usepackage{pdproc} 

\def\be{\begin{equation}}
\def\ee{\end{equation}}
\begin{document}

  \textwidth 6.0in
  \textheight 8.6in
  \pagestyle{empty}
  \topmargin -0.25truein
  \oddsidemargin 0.30truein
  \evensidemargin 0.30truein
  \parindent=1.5pc
  \baselineskip=15pt
\renewcommand{\thefootnote}{\alph{footnote}}

\title{
SN1987A REVISITED AFTER 20 YEARS:  MAY THE SUPERNOVA BANG MORE THAN ONCE?}

\author{P.GALEOTTI}
 \address{
  Dipartimento di Fisica dell'Universit\`a di Torino e INFN Sez. di Torino\\
   {\rm e-mail: piero.galeotti@to.infn.it}}
    \author{G.PIZZELLA}
  \address{
 Dipartimento di Fisica, Universit\`a di Roma ``Tor Vergata''\\
 INFN Laboratori Nazionali di Frascati\\
  {\rm e-mail: guido.pizzella@lnf.infn.it}
}

\normalsize\baselineskip=15pt

\abstract{
The observations of supernova 1987A in underground detectors are revisited. It is shown that, while the LSD detector in the Mont Blanc Laboratory observed only one burst at  $2^h52^{min}36.8^{sec}$ U.T., the Kamiokande data show a possible second burst, in addition to the well known one at  $7^h35^{min}33.7^{sec}$ U.T. This second burst consists of a cluster of seven pulses, well above the energy threshold of the detector, observed during 6.2 seconds starting at  $7^h54^{min}22.2^{sec}$ U.T. Do these observations imply a long duration of the collapse?
}

\section{Introduction}
Supernova 1987A  was a unique event during our time and a neutrino burst was
detected at 7:35 hour U.T. of 23 February 1987 in underground detectors: Kamiokande
in Japan, IMB in United States and Baksan in Russia. This burst occurred a few hours
before the optical detection of Supernova 1987A  by naked eye in  the southern
hemisphere, which triggered the search of pulses in these underground detectors. In
addition, another neutrino burst was observed on real time at 2:52 hours U.T.  by the neutrino
detector (LSD) located inside the Mont Blanc laboratory. This event was communicated immediately (IAU Circular n. 4323 of 28 February 1987) after
the visual supernova information was made available, discussed soon after on March 2
during the "Rencontres de Physique de la Vall d'Aoste" and sent to a journal
\cite{lsd}, about one month later. The Mont Blanc event, which occurred about four and
half  hours before the Kamiokande one, appeared surprising because it did not fit
most $a$ $la$ $page$ theories, according to which a gravitational stellar collapse
must occurre in a very short time, i.e. of the order of a few seconds or even less. 

Soon after the Supernova occurrence, new theories were proposed suggesting
that, because of fragmentation of a fast rotating core, the collapse could have
lasted for a few hours\cite{derujula,galeo}, thus allowing both the Kamiokande and
the Mont Blanc neutrino events. More recently the \it collapsar \rm model has been
proposed by Imshennik\cite{ims} and a paper has been published by Imshennik and
Ryazhskaya\cite{olgaim}, where a detailed mechanism is developed, based on the idea
that the collapsing star breaks under rotation in various pieces. In  this way
emission of gravitational waves could occour for several hours, precisely about five
hours according to\cite{galeo}, while the light fragments spiral around the
collapsed massive, central body.

In a simple rough description, the \it collapsar \rm model begins with the star
spinning very rapidly, taking the form of a thin disk (pancake type). The
neutronization process takes place:  $\nu_e$ are emitted in the energy range 20-55 MeV, and
possibly gravitational waves are also emitted (because of the non-spherical
collapse). Then the star breaks and the lightest fragments orbit around the neutron
star transferring matter to it. This process may have a duration of a few hours. The
transfer continues until the neutron star, with spherical shape, having reached a
sufficient mass undergoes a final collapse, probably into a black hole. Neutrinos
$\nu_e$ are produced and, after interacting within the star, all the six neutrino
species, are emitted with energy 10-20 MeV.

Since the first neutrino  $\nu_e$ emission is detected with high efficiency in Fe, and, among
the active detectors at that time, only LSD contained 200 tons of iron, this can explain the first burst. The other
detectors may detect neutrinos $\nu_e$ with lower efficiency. At this stage
gravitational waves could be emitted. The second neutrino emission (all six species)
is observed by the various detectors with efficiency proportional to their
mass. No gravitational waves are expected, because of the spherical symmetry of the
final collapse.

In view of this new approach to the supernova phenomena, twenty years later, we
believe we must examine again the data recorded in underground detectors, in particular the LSD and the
Kamiokande ones. We have found that, at the time of SN1987A, in addition to the burst detected by LSD, more
than one burst could have been detected by Kamiokande.

\section{Neutrino burst in the Mont Blanc detector}
The data we consider for the LSD detector consist in a list of triggers, in time and
energy (MeV), mostly due to background. We use the data recorded on 23 February (a
total of 1027 triggers) to compute the average trigger-rate of $0.0119
\frac{pulse}{second}$.

For the search of possible trigger clusterings we have applied the following
algorithm: \\
$\bullet$ Read the data and save them in a counter until more than 10 seconds
separate 2 successive pulses.\\
$\bullet$ Eliminate the content of the counter if the number of accumulated
triggers is less than 4.\\
$\bullet$ Given the time covered by the cluster (the last trigger time minus the
first one) and the previously measured background trigger rate, we calculated the
Poisson probability to have such a cluster (\em a posteriori\rm).\\
$\bullet$ We eliminate the cluster if the calculated Poisson probability is larger than 0.01.

With this procedure we obtained just one cluster: five pulses with energy in the
range 5.2 to 7.9 MeV detected within 7.0 seconds beginning at
$2^h52^{min}36.8^{sec}$ U.T., as reported in several papers of the LSD collaboration
in 1987 \cite{lsd}. The imitation rate from the background is 1 event every 6.2
years.

\section{Neutrino bursts observed in Kamiokande}
The Kamiokande data consist in a similar list of times and Nhits, being Nhit the
number of photo-multipliers hitted in the trigger. The calibration gives\cite{kami2}
an energy of 10 MeV for Nhit=26 and the energy of 30 MeV for Nhit=73. The Kamiokande
collaboration has put a threshold at Nhit=20, corresponding roughly to an energy of 7.5 MeV.
 In total the list we received from the Kamiokande collaboration contains 1937 triggers detected during 23 February above Nhit=20
for a rate of $0.024  \frac{pulse}{second}$. 

The search of possible trigger
clusterings, made by using the same procedure adopted for the LSD data, shows two
clusters, the first one is the well known burst described by the Kamiokande collaboration\cite{kami1,kami2} made by 11 pulses during 12.4 s, with a very low imitation rate from the
background. The second one, observed about 20 minutes later starting at $7^h54^{min}22.6^{sec}$,
is made by 7 pulses in a time window of 6.2 s with energy ranging from 22 to 33
Nhits and an imitation rate from the background of 1 event every 669 years.

In the Table \ref{tavek} we give the list of the pulses constituting the second burst.
 \begin{table}[b]
\centering
\caption{List of the pulses for the Kamiokande second burst. In the last column we give the number of years necessary for obtaining the cluster by chance.}
\vskip 0.1 in
\begin{tabular}{||cccc|c|c|c||}
\hline
\hline
hour&min&sec&nhit&number&duration&prob\\
&&&&&[s]&[years]\\
\hline
\hline    
    7&  54& 22.26&   33&7&6.2&669\\
    7&  54& 24.11&   29&&&\\
    7&  54& 25.33&   28&&&\\
    7&  54& 25.34&   27&&&\\
    7&  54& 27.13&   22&&&\\
    7&  54& 28.37&   22&&&\\
    7&  54& 28.46&   22&&&\\
 \hline
 \hline
\end{tabular}
\label{tavek}
\end{table} 

Since muons have been removed by the list of data we received from the Kamiokande
collaboration, and since the possible effects of muons on the pulses of
the first cluster has been studied very carefully by the Kamiokande group, see
ref.\cite{kami2}, we believe improbable that the second cluster
of triggers be due to muons. Indeed, the Kamiokande group find this possibility extremely small and concluded that the first cluster of pulses is due to neutrinos. It is our opinion that if the
second cluster, discussed in this paper, had not escaped to the Kamiokande team,
they would have discussed it in the same fashion as they did for the first cluster,
perhaps proving the muons effect. One can find an indication of this second cluster
in the fig.4 of ref.\cite{kami2}, from which, however, one does not realize that the
cluster consisted in seven pulses in just six seconds and well above background.

We must also comment about the coincidence with the IMB detector. As well known,  IMB  has energy threshold above 20 Mev. Thus, while IMB did observe signals in coincidence with the first Kamiokande cluster made by several high energy pulses, it could not have observed clustered signals in coincidence with the second Kamiokande cluster, where the higher energy was of the order of less than 15 MeV.

\section{Correlation with the gravitational wave detectors}
The gravitational wave group in Rome was immediately informed by the Mont Blanc collaboration (Carlo Castagnoli, private communication) about the burst detected by the LSD experiment. In spite of the low sensitivity of a room-temperature antenna in operation at that time (GEOGRAV), we decided to study carefully  the data. We  found a weak correlation with that burst, with the signal from the GEOGRAV detector anticipating the LSD signal by 1.4 seconds. This result was presented at the La Thuile meeting\cite{lath1} on 3 March 1987 and also published\cite{sn1}.

On 7 March we learned about the Kamiokande observation of a large neutrino cluster occurring about four and half hours after the Mont Blanc neutrino burst. 
In spite of the difficulty due to the Kamiokande observation at a later time\cite{kami1,kami2}, coincident with observation made with the IMB experiment, we thought important to continue the study of the GEOGRAV data, since there was a great chance that no other galactic, visible Supernova would have occurred for many years (or even centuries). In addition, also Joe Weber in Maryland made observations with his room temperature detectors at the time of supernova 1987A, and these appeared to have some degree of correlation with GEOGRAV. Then, our key idea was to consider \bf all \rm the LSD pulses and to try to correlate \bf all the available data\rm, and not just those occurring at  $2^h52^m36^s$.

We found a very strong correlation\cite{lath2,sn2} of both the Rome and the Maryland detectors with the LSD neutrino detector\footnote{These experimental results, presented by an American, Italian and Russian collaboration, were not believed by a large part of the scientific community, who did their best to ban the publication. The question, whether the strong correlation found by the above collaboration was true, would have been easily solved if another group had asked to analyze the real experimental data.}, with a time delay of 1.1 seconds in LSD and lasting for a period of one or two hours centered at the LSD time. The time shift was obtained with 101 LSD triggers, during a period of two hours centered at the Mont Blanc time, only 0.3 seconds off from the result presented\cite{sn1} several months before, when we used only the burst of  five neutrinos.

Later on we obtained  the Kamiokande data  for the day 23 February, including both the time of the Kamiokande observed neutrino burst and that of  the Mont Blanc burst. The Kamiokande data, recorded in an experiment aimed to measure the proton lifetime, had  a time uncertainty of  $\pm 1~minute$, but the time could be adjusted by imposing a coincidence with the IMB event at 7:35 hours. The correction  was 7.8 seconds.
With this time correction, applying the same procedure used for the correlation with LSD, we found\cite{sn3,sn5} a very similar correlation with the Kamiokande data during a period of one or two hours near $2^h52^m36^s$, just as found independently for the LSD detector, provided  we adjusted the Kamiokande  time by 7.8 seconds. 

\section{Conclusion}
In conclusion, the LSD data show a single cluster of pulses, probably due to $\nu_e$ interactions in Fe while, in addition to the well known cluster of pulses, the Kamiokande data show a possible second cluster, about 20 minutes later, with very small probability to be accidental and well above background. Forgetting theoretical prejudices and on the basis of all these observations without neglecting part of the data, we suggest a scenario with a long duration of the collapse. The LSD, Kamiokande, IMB and Baksan data may just be the \it top of an iceberg\rm, an activity that lasted for a few hours. 

We recall that the signals recorded by the gravitational wave detectors occur only during the first phase of the collapse, in a period that includes the $2^h52^m36^s$ LSD burst. These signals, observed by the Rome and by the Maryland detectors, show a strong correlations not only with the LSD data, but also and independently with the Kamiokande data, provided the time of all triggers is shifted in order to have a correlation with IMB.

We are aware that a problem arises if  the signals of the gravitational wave detectors are interpreted as due to gravitational waves, because their amplitude is very large.

\section{Acknowledgment}
We thank the Kamiokande group for supplying their data and Oscar Saavedra and Olga Ryashskaya for useful discussions.

\end{document}